\documentclass[sigconf]{acmart}
\usepackage{algorithm}
\usepackage[noend]{algpseudocode}
\usepackage{graphicx}
\usepackage{graphbox}
\usepackage{booktabs} 
\usepackage{subcaption}
\usepackage{tikz}
\usepackage{array}
\usepackage{multirow}
\usepackage[utf8]{inputenc}
\usepackage{amsmath,amssymb}
\def\BibTeX{{\rm B\kern-.05em{\sc i\kern-.025em b}\kern-.08emT\kern-.1667em\lower.7ex\hbox{E}\kern-.125emX}}

\begin{document}

\title{Identifying and Analyzing Cryptocurrency Manipulations in Social Media}

\author{Mehrnoosh Mirtaheri}
\affiliation{%
  \institution{USC Information Sciences Institute }
}
\email{mehrnoom@usc.edu}

\author{Sami Abu-El-Haija}
\affiliation{%
  \institution{USC Information Sciences Institute }
}
\email{haija@isi.edu}

\author{Fred Morstatter}
\affiliation{%
  \institution{USC Information Sciences Institute}
}
\email{fredmors@isi.edu}

\author{Greg Ver Steeg}
\affiliation{%
  \institution{USC Information Sciences Institute}
}
\email{gregv@isi.edu}

\author{Aram Galstyan}
\affiliation{%
  \institution{USC Information Sciences Institute}
}
\email{galstyan@isi.edu}
 
\renewcommand{\shortauthors}{Mirtaheri, et al.}

\begin{abstract}
Interest surrounding cryptocurrencies, digital or virtual currencies that are used as a medium for financial transactions, has grown tremendously in recent years. The anonymity surrounding these currencies makes investors particularly susceptible to fraud---such as ``pump and dump'' scams---where the goal is to artificially inflate the perceived worth of a currency, luring victims into investing before the scammers can sell their holdings. Because of the speed and relative anonymity offered by social platforms such as Twitter and Telegram, social media has become a preferred platform for scammers who wish to spread false hype about the cryptocurrency they are trying to pump. In this work we propose and evaluate a computational approach that can automatically identify pump and dump scams as they unfold by combining information across social media platforms. We also develop a multi-modal approach for predicting whether a particular pump attempt will succeed or not. Finally, we analyze the prevalence of bots in cryptocurrency related tweets, and observe a significant significant presence of bots during the pump attempts.     
\end{abstract}

%
%


%
%
\maketitle

\section{Introduction}
The inception of blockchain technology \cite{blockchain} gave birth to the popular cryptocurrency Bitcoin (symbol \texttt{BTC}). Since then, thousands of cryptocurrencies have emerged, and their hype has caused massive price swings on the trading markets.
In December 2017, \texttt{BTC} quadrupled in market value in just over a month, then within a few days started a gradual decline until it reached half of its peak value. These price changes allowed some investors to realize huge profits, contributing to the allure of cryptocurrencies.
Even though most investments are made in relatively established cryptocurrencies, including Bitcoin (\texttt{BTC}) and Ethereum (\texttt{ETH}), there are thousands of other smaller cryptocurrencies. These currencies are prime targets for manipulation by \textit{scammers}, as evidenced by the proliferation of pump and dump schemes.

\begin{figure}
\centering
\begin{tikzpicture}
\node{\includegraphics[height=2.5in]{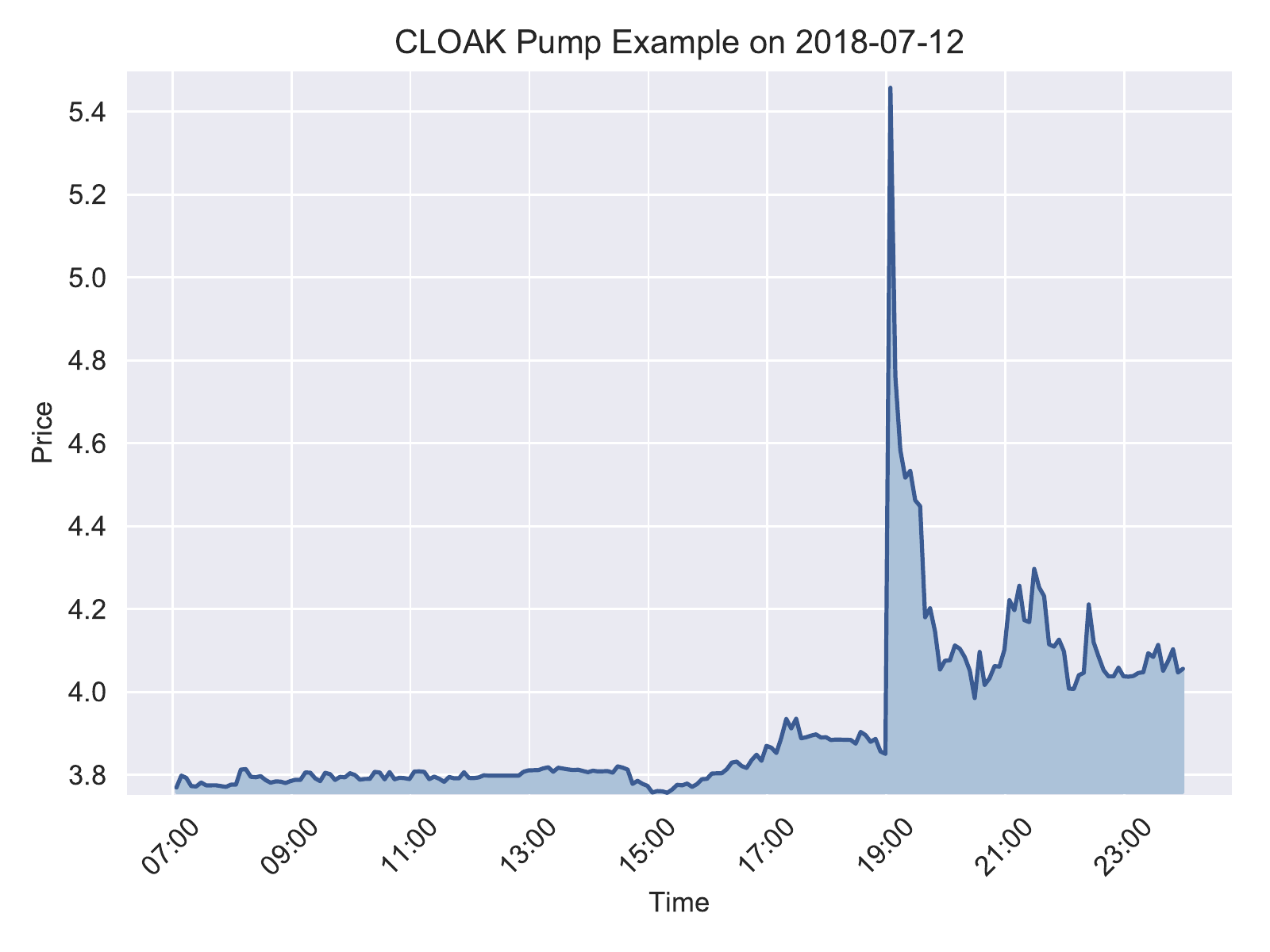}};
\node[](peakpoint) at (1.6cm,2.6cm){};
\node[](word) at (2.5cm,2cm){\textcolor{red}{\shortstack{scammers \\start sell here}}};
\draw[->,red] (word) to [bend right] (peakpoint);
\node{\includegraphics[width=3cm]{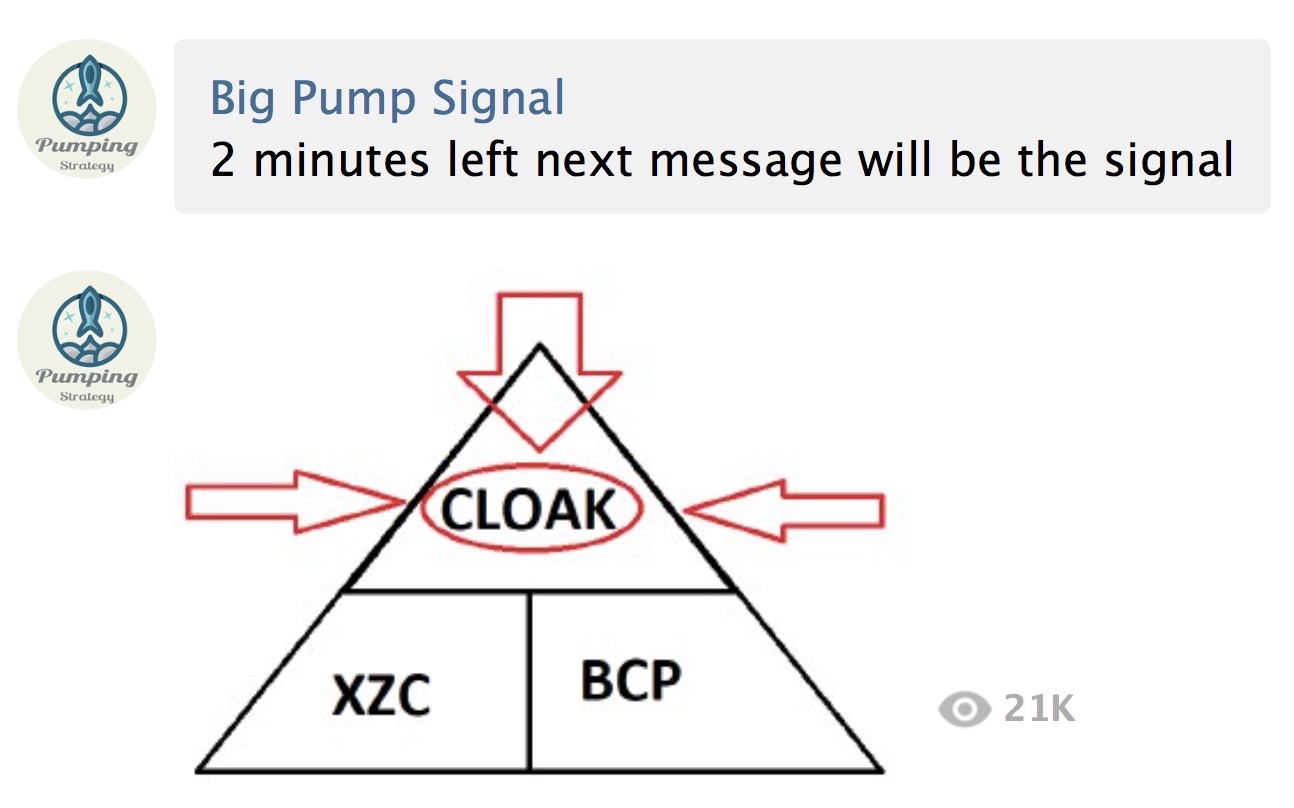}};
\draw[->,red](0.9, -0.9) to [bend left] (1.65, -1.8);
\end{tikzpicture}
\caption{Pump announcement by admin of a Telegram Channel ``Big Pump Signal,'' overlayed on the market values of \texttt{\$CLOAK}. The announcement precedes the price swing.}
\label{fig:pump-ts-exp}
\end{figure}
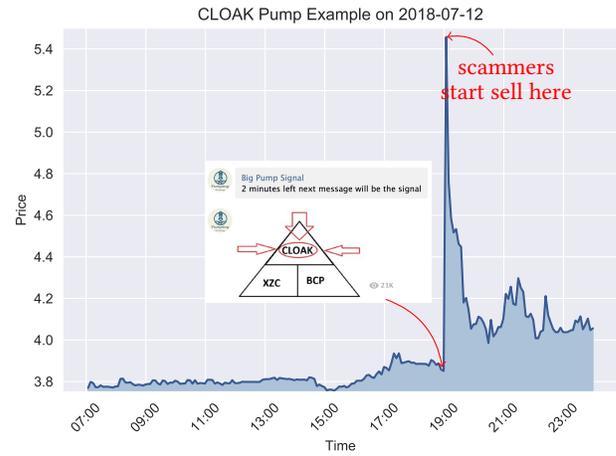


Pump and dump schemes are those in which a security price inflates due to deliberately deceptive activities. Those fraudulent schemes originated in the early days of the stock market and are now growing rapidly in the cryptocurrency market. The fact that the Commodity Futures Trading Commission (CTFC) and U.S. Securities and Exchange Commission (SEC) have issued several warnings \cite{pumpdumpgov} against cryptocurrency pump and dump schemes highlights the severity of the threat. 

Although in the early days of cryptocurrencies, pump and dump schemes were taking place by marketing teams in ICOs \cite{pumpdumpico} (Initial Coin Offerings), they are taking different forms nowadays. Pump and dump schemes have three major components: (1) a group of scammers; (2) a private or semi-private communication medium where scammers coordinate their illicit activities; and (3) a social media platform for conducting orchestrated campaigns to hype a given coin. In a typical scenario, scammers create groups on platforms such as Telegram or Reddit to coordinate group purchases of a particular cryptocurrency, while creating false hype around it by making public posts (i.e., \textit{pump}) on social media platforms such as Twitter. Normal traders, who only see the rise in the price and are unaware of malicious activity, might buy the coin hoping to anticipate the next trend, thus boosting the price even further. Once a certain price target is met, the scammers start to sell (i.e., \textit{dump}) their holdings, leading to a precipitous drop in the price. We illustrate the process with an example mentioned in a Wall Street Journal article\footnote{\url{https://www.wsj.com/graphics/cryptocurrency-schemes-generate-big-coin/}} that we also found in our collected data. In this example, the fraudsters coordinated their activities using a public group ``Big Pump Signal'' on the Telegram messaging app, which has more than $60,000$ members. As shown in Figure \ref{fig:pump-ts-exp}, after the  the coin CLOAK posted on the group at 7:00 PM GMT, the price inflated immediately after the pump message and dropped shortly afterwards.

As cryptocurrency trading attracts more public attention, it becomes extremely important to be able to detect such fraudulent activities and inform potentially susceptible people before they become victims of these crimes. Toward this end, we study the extent to which pump and dump groups on online forums, such as Telegram, are accompanied and correlated with suspicious activity on Twitter and cryptocurrency price movements. In addition, we quantify the ability of machine learning models to predict pump and dump schemes from our data sources. 
In particular, we propose and address the following two predictive tasks:
\begin{enumerate}
    \item Predict an unfolding pump operation/advertisement campaign happening on Telegram.
    \item Predict whether a detected operation will succeed, i.e., will the target price mentioned in the Telegram message be reached by the market shortly after the announcement?
\end{enumerate}

Although closely monitoring covert communication of fraudulent users on Telegram enables us to detect pump and dump schemes, in many realistic scenarios we might not have access to such communication. For instance, some scammers are using private channels with restricted access and membership fees, or communicate on different platforms altogether. Furthermore, most pump and dump events happen in a very short time after a coin is announced on Telegram. Thus, we examine the possibility of detecting pump and dump schemes {\em without} relying on availability of the Telegram data. Remarkably, we demonstrate that it is indeed possible to detect pump and dump events by leveraging only publicly available information, such as Twitter and historical market data.

Our main contributions are as follows:
\begin{enumerate}
    \item We propose a multi-modal approach to monitor potentially malicious activities in cryptocurrency trading by combining data from three distinct sources: $(i.)$ Real-time market data on crytpocurrency trading including both price and volume information; $(ii.)$ Twitter data of {\em cashtag} mentions for cryptocurrencies; and $(iii.)$ Telegram data that contains potential mentions of and instructions for pump and dump activities. 
    
    \item We identify pump and dump operations on Telegram messages by manually  labeling a fraction of those messages as ``\textit{pump}'' versus ``\textit{not-pump}'', and then building a classifier to label the remaining messages. Our approach is efficient for dealing with huge numbers of unlabeled messages mentioning cryptocurrencies and achieves a high precision (Section \ref{sec:pump-detection}).
    
     \item In Section \ref{sec:pump-predict}, we explore the possibility of forecasting specific pump and dump activities, as quantified on the two classification tasks, based on different combinations of features extracted from the above-listed data sources. Our results indicate that it is indeed possible to forecast such events with reasonable accuracy.
     
    \item We study the efficacy of pump and dump operations on cryptocurrency price movements and Twitter activity in Section \ref{sec:pump-efficacy} and investigate the prevalence of Twitter bots in cryptocurrency-related tweets, especially during the alleged pump and dump attacks, and observe that the majority of highly active users are bots. (Section \ref{sec:bots}) 

    \item We release\footnote{Available at https://github.com/Mehrnoom/Cryptocurrency-Pump-Dump} a comprehensive dataset containing coins, time-stamps, pump messages, and indications of whether or not the pumps were successful---together with features we extract from Twitter.
\end{enumerate}

\section{Data Description}
\label{sec:data}
In this section we describe the datasets used in our study, explain the data collection process, and provide basic statistics of the data. 

\begin{figure}[t]
\centering
\includegraphics[width=0.4\textwidth]{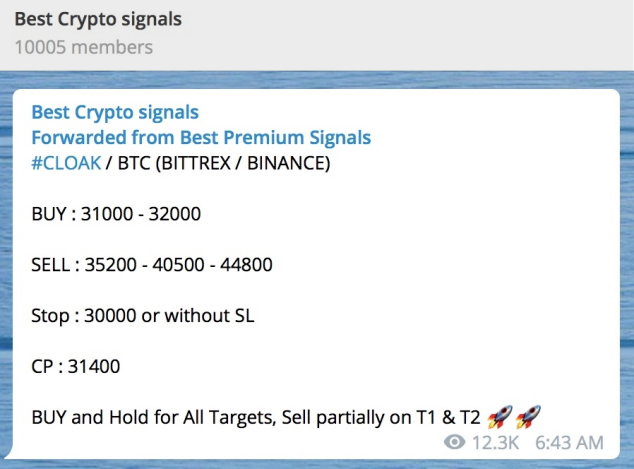}
\label{fig:tlgrm-msg}
\caption{The anatomy of a common pump announcement, containing a ``BUY'' price and multiple target ``SELL'' prices. ``CP'' and ``Stop'' refer to the current coin price and stop loss respectively.}
\end{figure}

\subsection{Telegram Data}
\label{sec:telegramdatasource}
In the context of the cryptocurrency market, scammers coordinate within groups to inflate the market value for a coin using  social media platforms. In particular, the messaging platform Telegram is widely used for sharing cryptocurrency-related information, including pump announcements. The reason for Telegram's popularity among scammers is that it provides anonymity for the users. For example, a Telegram channel consists of an anonymous admin and a set of members; however, the only person who can post to the channel is the admin who also is the only one who can see the list of members, while his/her identity is anonymous to the members.

We implemented a crawler using the Telegram API\footnote{https://telethon.readthedocs.io/en/latest/} to collect data from Telegram channels. The crawling process starts with a set of a few initial channel IDs, and then we extend the list by extracting the other channels' hyperlinks advertised in the seed channels and add those channels to our set. We continue to snowball out as new channel IDs appear in the Telegram channels. Due to the nature of how private channels are joined (the password to a private channel can be passed via its URL), it is entirely possible that we would crawl private channels if such URLs are posted in a public channel. We make no distinction between the two classes in our experiment.

Table \ref{tab:tlgrm-stats} shows the statistics of the Telegram data. We extracted all the messages containing at least one coin from our coin list including all the cryptocurrencies provided by \texttt{CoinMarketCap.com}, which resulted in 195,576 messages. The Telegram channels in the table are categorized by their size (number of members) because channels with more members are more likely to contribute to a pump event. 
\begin{table}[!h]
\centering
\caption{Telegram data statistics. Top: Histogram of number of channels grouped by their size (i.e., members in each channel). Middle: Summary statistics. Bottom: Number of messages containing the coins in our coin list.}
  \label{tab:tlgrm-stats}
  \begin{tabular}{lc}
    \toprule
    Channel Size(S) & Count \\
    \hline
     $ S \in [0, 100]$&$51$\\
     $ S \in (100, 1000]$ & $159$\\
     $ S \in (1000, 10000]$ & $50$\\
     $ S \in (10000, \infty)$ & $63$\\
     \hline
     Number of channels & $423$\\
     Average channel size & $6,084$ \\
     Median channel size& $1,005$\\
     \hline
     Total number of messages& $195,576$\\
  \bottomrule
\end{tabular}
\end{table}
\subsection{Twitter Data}
\label{sec:twitterdatasource}
One way of promoting finance-related information on social media, especially Twitter, is to use \textit{cashtags}, which are ticker symbols of stocks or cryptocorrencies prefixed by \texttt{\$}, e.g.,  \texttt{\$BTC} is the appropriate cashtag for Bitcoin. Using the Twitter streaming API,\footnote{https://developer.twitter.com/en/docs/tweets/filter-realtime/overview.html} we implemented a system that tracks all the cryptocurrencies provided by \texttt{CoinMarketCap.com}, including 1,600 cashtags. We began the data collection process on March 15, 2018. However, cryptocurrencies had received a considerable amount of attention prior, toward the end of the 2017, due to growth in the Bitcoin price which paved the path for fraudulent activities.
For better coverage of the potentially fraudulent events occurring before March 15, 2018, when we began our data collection, we also purchased tweets from September 1, 2017, to March 31, 2018. 

\begin{figure}[!h]
\centering
\includegraphics[width=0.4\textwidth]{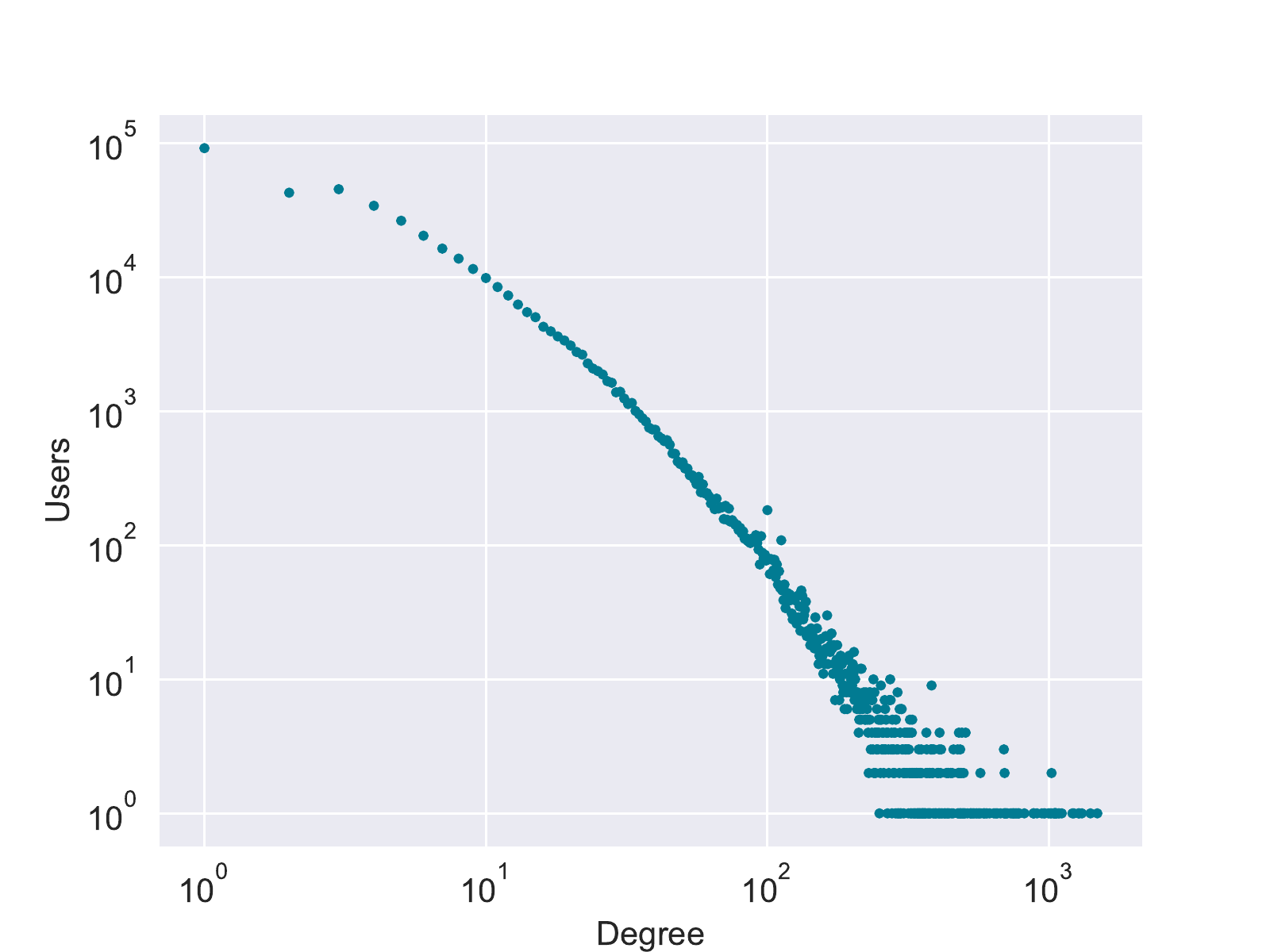}
\caption{Users' degree distribution, showing most users are interested in only a few cashtags, while some users discuss a large number of cashtags.}
\label{fig:deg-dist-user}
\end{figure}

The resulting dataset includes 30,760,831 tweets and 3,708,176 users in total from September 1, 2017, to August 31, 2018. Figure \ref{fig:deg-dist-user} also shows the distributions of the number of users per cashtag. The distribution is heavy-tailed, which suggests that many users are interested in only a few cashtags, while a small number of users tweet about many cashtags. 


\subsection{Cryptocurrency Market Data}
This dataset consists of the time-series of market values for many cryptocurrencies.
We developed a crawler to collect data from \texttt{CoinMarketCap.com}. Instead of using end-of-the-day historical data, we chose to use data with a  five-minute granularity because pump and dump schemes usually happen within a very short period. 
The dataset includes all the coins listed on the website at the time we started the collection process.

\section{Pump and Dump Activities  on Telegram}
\label{sec:pump-detection}
As mentioned above, Telegram is a popular choice for scammers to organize and coordinate pump and dump operations. To analyze such activities, let us define the following two notions: 
\begin{enumerate}
    \item \textbf{Pump attempt}: The act of targeting a coin on Telegram by posting a \textit{pump} message mentioning the coin as a ``pump attempt.'' In Section \ref{sec:telegramtextclassifier} we describe our approach to detect pump messages.
    \item \textbf{Successful pump attempt}: A pump attempt is \textit{successful} if the actual price approaches 
    the target price  within a time window after the first pump message has been posted. 
\end{enumerate}
We next describe a simple method for detecting individual pump attempts and assessing whether those attempts were successful. The pump attempts (either successful or not) are used as ground truth for building and evaluating predictive models proposed in Section~\ref{sec:pump-predict}.

\subsection{Pump Message Identification on Telegram}
\label{sec:telegramtextclassifier}
Most of the Telegram messages in our dataset are about cryptocurrency-related news, advice, and advertisements that are not relevant in the context of predicting pump and dump activities. Only a small fraction of those messages contain actual pump announcements. As shown in Table \ref{tab:tlgrm-stats}, the number of cryptocurrency-related messages is  large, which would make it prohibitively expensive to manually label them as pump-relevant or not. Fortunately, however, most of the pump announcements follow specific patterns, or redundancies, which we are able to detect with machine learning techniques.

Text is the most common format used by Telegram channels for broadcasting pump events, although some channels embed the coin name in an image to prevent trading bot activities (Figure \ref{fig:pump-ts-exp}). A pump text message includes the name of the coin, the price to buy the coin, and one or more desirable target prices to achieve. 

We leverage this specific common pattern to extract pump-related messages using a weakly supervised approach. Toward this goal, we  labeled 1,557 messages in total as pump/not-pump, coming from 15 channels. To avoid bias toward a specific coin, we replace all the cryptocurrency symbols based on  whether they are  OOV (object out of vocabulary). We represent each post as a TF-IDF vector. For a given post, an entry in its TF-IDF vector corresponds to the frequency of a token appearing in the post (TF) divided by the number of posts in which the token appeared (IDF). In general, the size of the vector is equal to the size of the vocabulary of the entire corpus. We can use word $n$-grams (a sequence of $n$ words) to construct the vocabulary.

We train a linear SVM with an SGD optimizer; this achieves an accuracy of $87\%$ and a precision of $89\%$. The optimal parameters for linear classifier and TF-IDF tokenizer are obtained by cross-validation. The best result is achieved when we use both unigrams and bigrams, $\max_{DF} =0.5, \min_{DF}= 0.01$, and $L_2$ penalty. The classifier scores are included in Table \ref{tab:tlgrm-cls}. Using the trained model, we then label the entire messages as pump/not-pump. From each message, we extract the coins mentioned in the message and the message timestamp. We call the (coin, timestamp) pair a pump attempt. We aggregate multiple timestamps into the earliest one, if they appear within a 3-hour window. We also remove the messages that mention many coins, as a pump message usually targets only a small number of coins. The statistics of the final set of messages and pump attempts are given in Table \ref{tab:pump-signal-stats}. 

\begin{table}
\centering
    \caption{Telegram pump detection performance}
  \label{tab:tlgrm-cls}
  \begin{tabular}{ccccc}
    \toprule
    \textbf{Base Rate}&\textbf{Accuracy}&\textbf{Precision}&\textbf{Recall}&\textbf{F1}\\
    \midrule
    0.603&0.879& 0.895& 0.908& 0.901\\
  \bottomrule
\end{tabular}
\end{table}

\begin{table}
\centering
    \caption{Pump attempts statistics}
  \label{tab:pump-signal-stats}
  \begin{tabular}{cc}
    \toprule
   Pump Attempts & $62,850$\\
    Pump Messages & $47,992$\\
    Channels & $209$\\
    Coins & $543$\\
  \bottomrule
\end{tabular}
\end{table}

\subsection{Signatures of Pump and Dump Activities on Market and Social Data}
\label{sec:pump-efficacy}
Having established the prevalence of pump and dump operations on Telegram, we now analyze the effectiveness of those operations, by juxtaposing pump messages with crytocurrency market data and social signals collected from Twitter. Specifically, we focus on the overall effect of pump attempts on cryptocurrency prices and determine whether there is an indication of concurrent fraud activity on Twitter.

Let $S_i = (c_i, t_i)$ be a pump attempt, with $t_i$ the time of the attempt, and $c_i$ the target coin (in the next section, we explain in detail our approach for detecting pump attempts). For every pump attempt $S_i=(c_i, t_i)$ we extract two time series segments: The price and tweet volume of coin $c_i$, denoted as $\Tilde{P}^{(c_i)}_{t_i-w:t_i+w}$ and $\Tilde{V}^{(c_i)}_{t_i-w:t_i+w}$ respectively, where $w$ is a time window equal to three days. Each segment is normalized between 0 and 1 by a minmax normalization, transforming each point $x$ to $\Tilde{x}= (x-\min)/(\max-\min)$.
We calculate $Q^{price}$ and $Q^{twitter}$ as follows: 
\[Q^{price} = \frac{1}{n}\sum_i^n P^{(c_i)}_t \textrm{ \ \ and \ \ } Q^{twitter} = \frac{1}{n}\sum_i^n V^{(c_i)}_t,\]
Note that $i \in [1, n]$ means that we have considered all the pump attempts. For each coin, we also select a set of timestamps uniformly at random, with the size equal to the number of pump attempts targeting the coin. In the same manner, we make two aggregated timeseries for the random timestamps. 

\begin{figure*}[!h]
\centering
\begin{subfigure}[t]{0.45\textwidth}
\centering
\includegraphics[width=\textwidth]{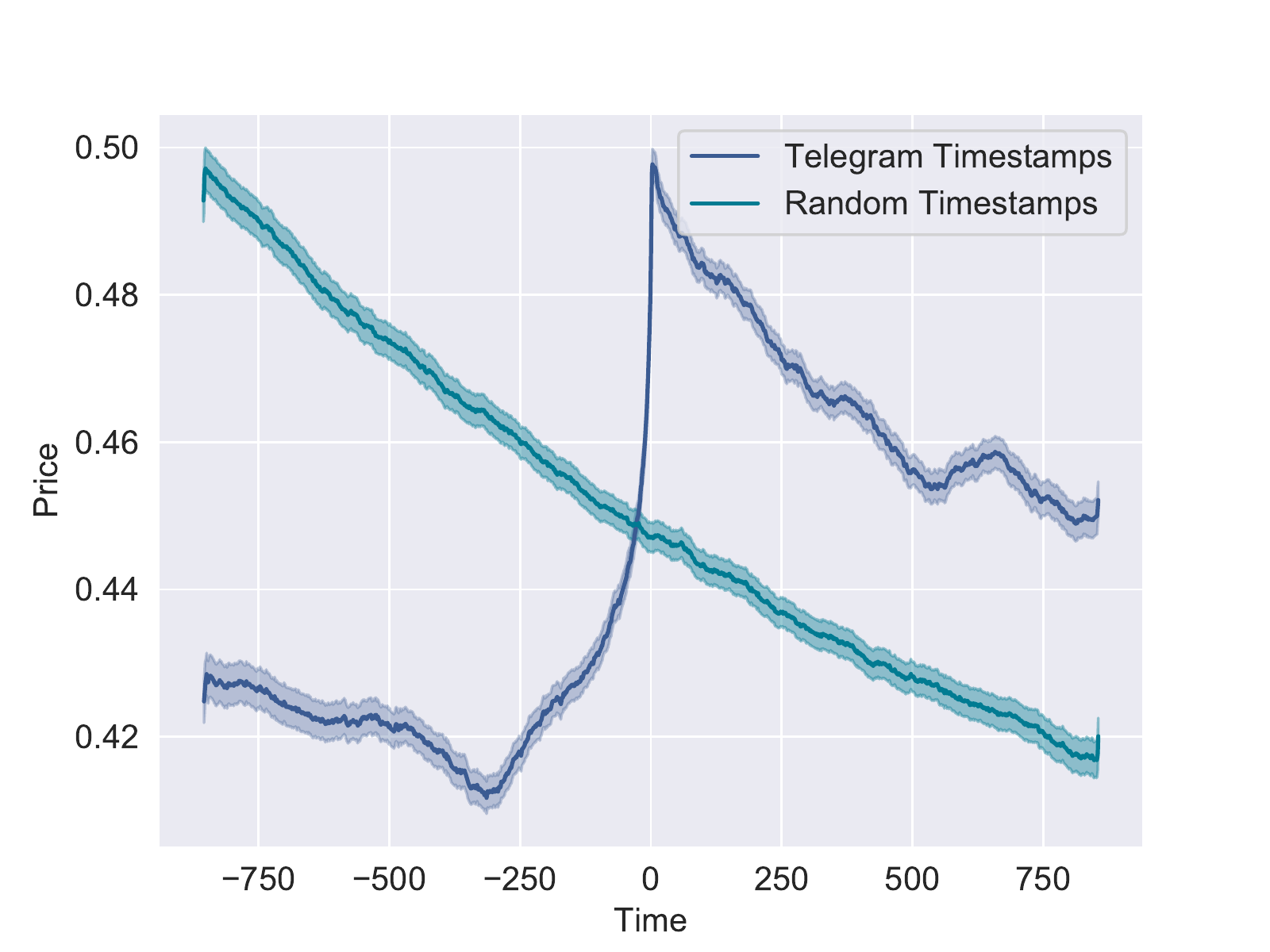}
\caption{Aggregated Price}
\label{fig:aggregate-ts-price}
\end{subfigure}
\begin{subfigure}[t]{0.45\textwidth}
\centering
\includegraphics[width=\textwidth]{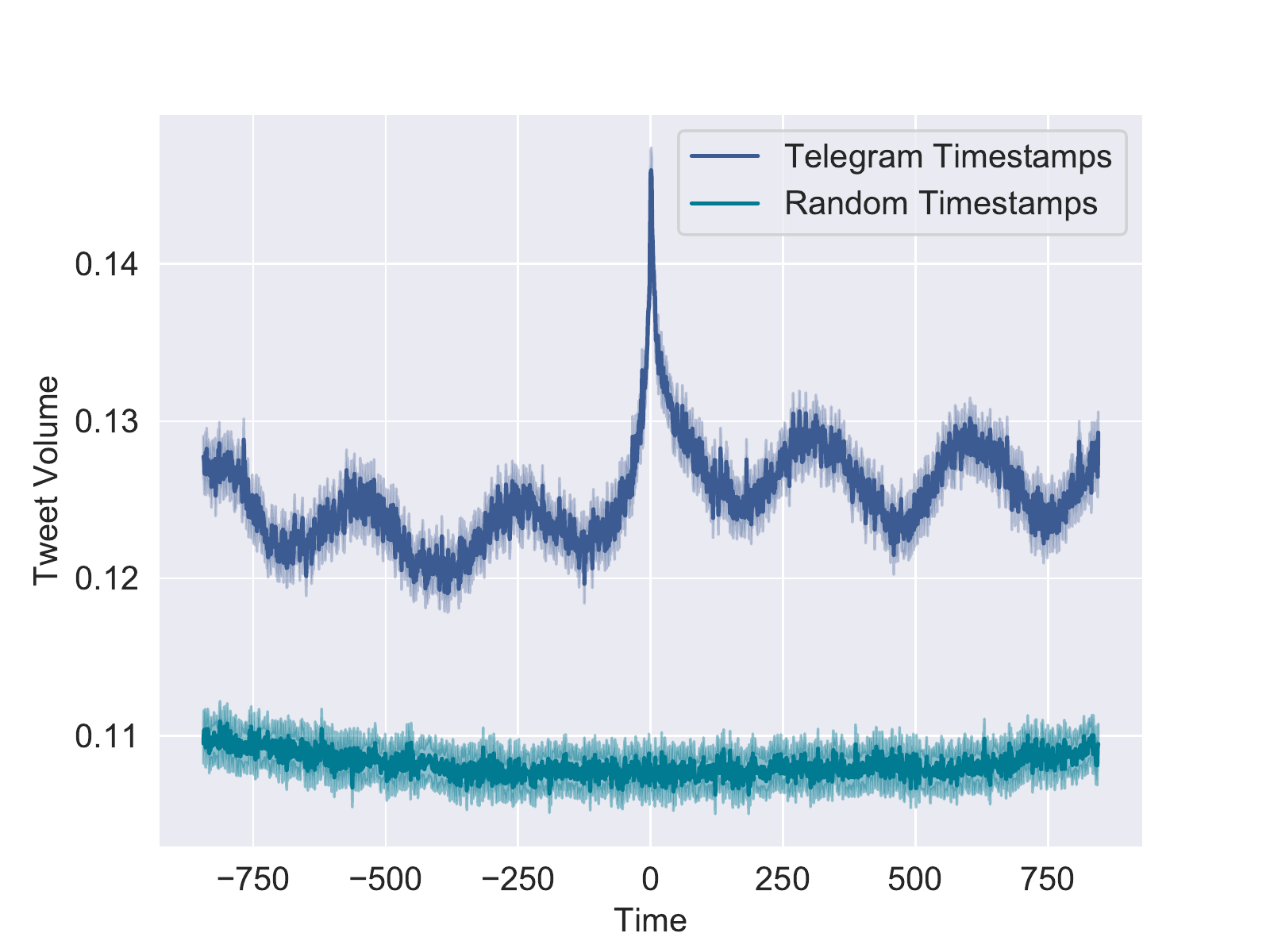}
\caption{Aggregated Tweet Volume}
\label{fig:aggregate-ts-twitter}
\end{subfigure}
\caption{Price (a) and Twitter (b) time series segments of cryptocurrencies separated by whether they are mentioned in a Telegram pump and dump channel. Segments start 3 hours before and end 3 hours after the mention in a Telegram channel. Timestamps are selected uniformly at random.}
\label{fig:aggregate}
\end{figure*}

Figure \ref{fig:aggregate-ts-price} depicts an average normalized price of each coin centered around the pump timestamps ($Q^{price}$) and random timestamps. This figure shows a pattern of spikes occurring within one hour of a pump message, followed by a general downward trend.
A significant increase in tweet volume is also observable in Figure \ref{fig:aggregate-ts-twitter}, which shows the average tweet volume of each coin around the pump timestamps ($Q^{twitter}$) and random timestamp. The seasonality pattern present in the Telegram timestamps, and not in the random timestamps, suggests that most of the pump attempts happen around a specific time of day. 

\subsubsection*{Successful Pump Attempts}
\begin{figure}[t]
\centering
\includegraphics[width=0.45\textwidth]{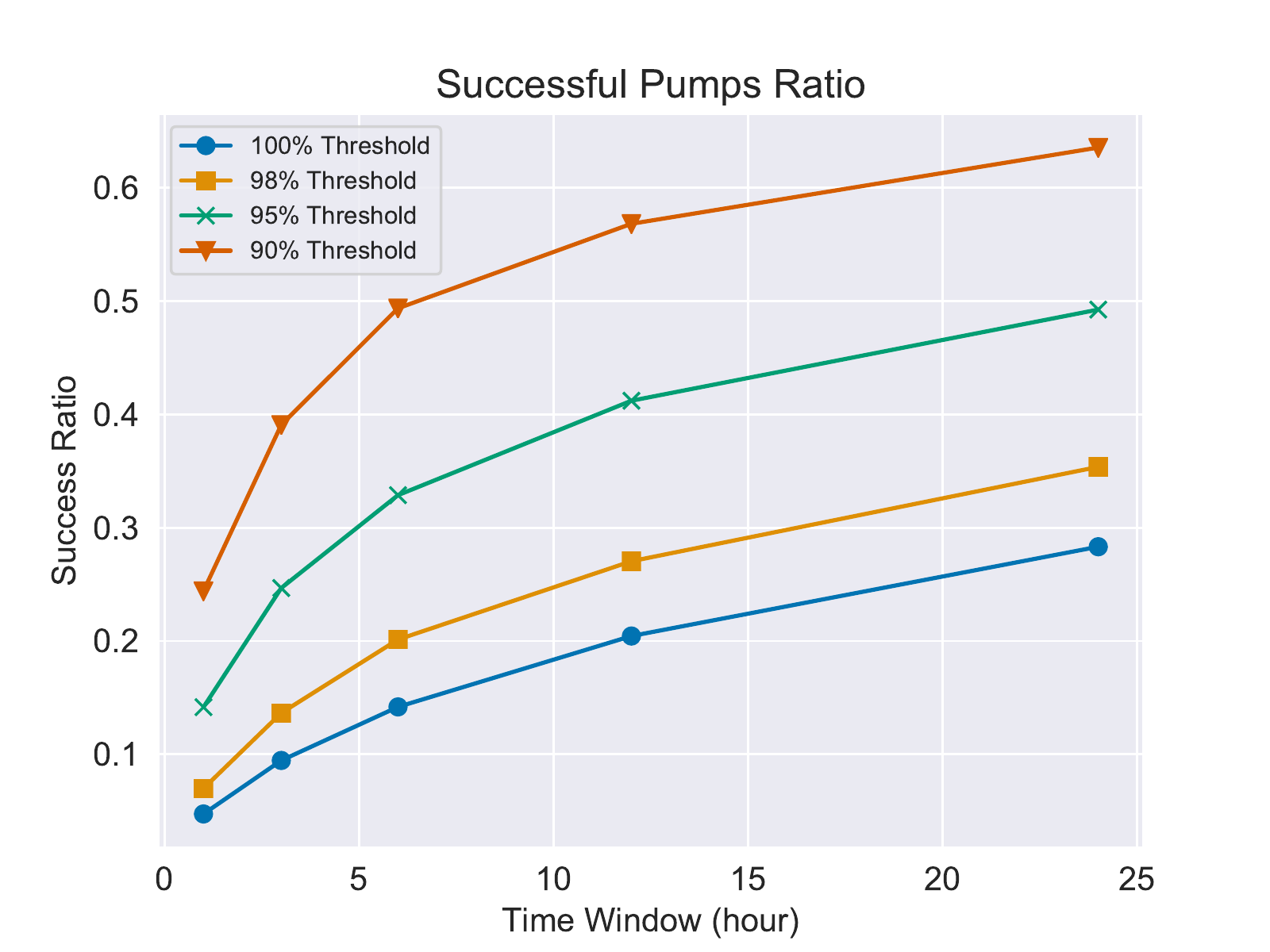}
\caption{Ratio of successful pumps for different thresholds and time windows. ``100\% threshold '' indicates the coin price beats the extracted target price. ``$x\%$  threshold'' is when the coin price either beats the price or reaches the $x\%$ of the extracted price.}
\label{fig:success-pump}
\end{figure}

Although Figure~\ref{fig:aggregate-ts-price} indicates a price spike for the aggregated data, it is reasonable to expect that not all the pump messages are actually followed by a spike in the coin price. Furthermore, even if there is a spike, it might fall short of meeting the ``target" price, thus resulting in \textit{failed} pump. We used a simple rule-based approach to extract the ``buy" and ``target" prices from all the messages, and augmented this information with the coin price data to decide whether a given pump attempt was successful or not.   
Figure \ref{fig:success-pump} shows the percentage of successful pump messages for different thresholds and time windows. Fewer than $5\%$ of the pump messages meet the most strict conditions, meaning that the coin price reaches a higher price than the extracted target price, and within an hour of the pump message.

\section{Pump Attempts Prediction}
\label{sec:pump-predict}
In this section we study the feasibility of predicting pump and dump events from the social media and market data only, without relying on the availability of Telegram messages~\footnote{We use Telegram messages only as ground truth for evaluating our predictive models, but not as input to those models.}. Specifically, we focus on the following two classification tasks:
\begin{itemize}
    \item Task I: Detect whether there is an unfolding pump operation/advertisement campaign happening on Telegram, by considering social signals {\em only} from Twitter, and historical market data. 
    \item Task II: Given a pump message on Telegram for a specific coin, predict whether the operation will succeed (i.e., will the target price, set by the scammers, be met within 6 hours of the message is posted on Telegram).
\end{itemize}
Both of these tasks are cast as binary classification problems. The feature vector for each record is extracted at a specific timestamp: specifically, it contains data from 6 hours prior to the timestamp. The features of the timestamps are explained next.

\subsection{Features}
\label{sec:features}
Table \ref{tab:features} explains all the features extracted from our data sources that are used for the two prediction tasks. Graph features include: (i.) PageRank score; (ii.) CorEx user embeddings; and (iii. ) user connected components. These are extracted from the coin-coin network, the pump-user network, and the user-user network, respectively. These networks are undirected and temporal, where the presence (or absence) of an edge depends on the time range in which the networks are constructed. Each network is explained in its corresponding feature section. 

\begin{table*}
\centering
\caption{Description of the features used in the pump predictions and user clustering.}
\label{tab:features}
\begin{tabular}{c| c p{12 cm}}
\toprule
 \textbf{Feature type} &  & \textbf{Description}\\
\midrule
\multirow{6}{*}{Twitter Features} & $\bullet$ & Number of tweets mentioned the cashtag in the period $[t, t-w]$\\
    &$\bullet$ & Number of unique users mentioned the cashtag in the period $[t, t-w]$\\
    &$\bullet$ & Average sentiment of all the tweets mentioning the cashtag in the period $[t, t - w]$ calculated using \cite{gilbert2014vader}\\
    &$\bullet$ &  PageRank score \cite{pagerank} of a coin in the Coin-Coin graph created at time $t$ \\
    &$\bullet$ & Twitter User Connected Components \\
    &$\bullet$ & CorEx user embedding \\
\midrule
\multirow{3}{*}{Economic Features} & $\bullet$ & Coin market cap and market cap percentage change at $h$ hour before the pump where $h \in \{1, 2, \dots , 12\}$ \\
    &$\bullet$ & Coin volume and volume percentage change at $h$ hour before the pump where $h \in \{1, 2, \dots , 12\}$\\
    &$\bullet$ & Coin price and price percentage change in BTC unit at $h$ hour before the pump where $h \in \{1, 2, \dots , 12\}$\\
    &$\bullet$ & Target price percentage difference with coin price at $h$ hour before the pump where $h \in \{1, 2, \dots , 12\}$\\
\bottomrule
\end{tabular}

\end{table*}

\subsubsection*{Economic Features} Include (i) Coin Market Cap, (ii) Volume, and (iii) Price, in $h \in \{1, 2, \dots, w\}$ hours before timestamp $t$. For each of these three, we also include percentage change. For example, price percentage change at hour $h$ is $\frac{\textrm{price}_{h+1} - \textrm{price}_h}{\textrm{price}_h}$.

\subsubsection*{Target Price Features}This feature will only be used in the second classification task (whether the pump succeeds, introduced in Section \ref{sec:pump-predict}). As explained in Section \ref{sec:pump-detection}, the definition of success depends on the target price mentioned in a pump message. For a given pump message with a mentioned target price $x$, we include $\frac{x - \textrm{price}_h}{x}$ for $h \in \{1, 2, \dots, w\}$ hours before the pump message timestamp.

\subsubsection*{Twitter Statistics.} For a given timestamp $t$, the features extracted from time period $[t, t - w]$ include (i.)  the number of tweets mentioned the cashtag; (ii.) the number of unique users who mentioned the cashtag in a tweet; and (iii.) the average sentiment of all the tweets mentioning the cashtag calculated using \cite{gilbert2014vader}.

\subsubsection*{PageRank Score.} We calculate the PageRank score~\cite{pagerank} of a coin in the \textbf{coin-coin} graph created at time $t$, in which the nodes correspond to the cashtags, connected by an edge if they are mentioned by the same user, within the period [t - w, t]. The edge weights correspond to the number of times two coins are co-mentioned by a user.  

\subsubsection*{CorEx User Embeddings.} 
\label{sec:corex-emb}
Consider a bipartite graph containing pump attempts and users as nodes. The edge weight between a pump attempt $S_i = (c_i, t_i)$ and user $u_j$ is equal to the number of times $u_j$ mentions $c_i$ in a tweet within the period $[t-w, t]$. Here we chose $w= 6$ hours. We call this graph \textbf{pump-user network} and denote $\mathbf{B} \in \mathbb{R}^{|S|\times|U|}$ as the affiliation matrix of this bipartite graph, where $S$ and $U$ are the set of pump attempts and users respectively. $\mathbf{B_{ij}}$ is equal to the weight of the edge connecting $S_i$ and $u_j$. In fact, matrix $\mathbf{B}$ represents each pump attempt as a $|U|$-dimensional vector of user activities, meaning that each user is considered as a variable. We want to cluster users if their activity are correlated. 
Total Correlation Explanation (CorEx) discovers a latent representation of complex data based on optimizing an information-theoretic approach. More specifically, given a set of $n\textrm{-dimensional}$ vectors $\mathbf{X} \in \mathbb{R}^{m\times n}$, CorEx aims to find latent variables $\mathbf{Y} = \mathbf{WX}$  that best describe the multivariate dependencies of $\mathbf{X}$ by minimizing $TC(\mathbf{X|Y}) + TC(\mathbf{Y})$. Here, $TC$ is ``total correlation'' or multivariate mutual information \cite{watanabe1960information}. 

We apply linear CorEx on $\mathbf{B}$, and find the best number of latent factors $k$ by plotting the sum of the total correlation for each latent variable against $k$. This value starts descreasing significantly when $k = 24$. The weight matrix $\mathbf{W} \in \mathbb{R}^{m \times k}$ obtained from applying linear CorEx is used as the embedding for the users. Later in Section  \ref{sec:bots} we further explore pump-user network and CorEx clusters for analysing bot activities.
\subsubsection*{User Connected Components} 

We would like to characterize groups of users that participate in coordinated spreading of ``pump" messages on Twitter. Given a coin $c$ and all the pump instances targeting $c$, $S_i = (c, t_i)$, a bipartite graph is built from the pump instances and users. $u_j$ is connected to $S_i$ if $u_j$ tweeted about $c$ within the time period $[t_i-6, t_i]$, measured in hours. From this bipartite graph, we then build a \textbf{user-user} network, in which users $i$ and $j$ are connected with a weighted edge that corresponds to the time the users co-tweet about $c$. After this process, the graph is very dense with usually $>80\%$ of nodes belonging to one connected component. We sparsify the graph by keeping the top-$k$ edges per user. Then we calculate the connected components from the graph, dropping connected components consisting of less than 25 users. Each user is now represented by the ID of its connected component. Users that do not correspond to the connected components are ignored in this feature representation. Manual inspection of the connected components showed that they are indeed meaningful, with the largest connected component usually corresponding to users who are likely involved in pump and dump operations.
Given a coin and a timestamp, we create a feature vector containing the counts of users tweeting about the coin, grouped by the connected components they belong to. Intuitively, these features represent how active a group (connected component) has been within a period from $t - w$ to $t$. We tried $k \in \{1, 2, 3\}$, and the final result does not change much. However, higher values produced just a handful of sizable connected components.

\subsection{Classification Tasks}

In this section we explain the experimental setting and prediction tasks design. 

\subsubsection*{Task I: Predicting Pump Attempts.} We propose a binary classification task for predicting Telegram pump messages that will happen in the future, from Twitter activity. In our setup, we use the timestamps of Telegram messages labeled as pump attempts by our classifier as positives. We use an equal number of random timestamps as negatives.

\noindent\textbf{Task II: Will the Pump Succeed?}
The positives of this task are a subset of the positives in Task I---namely, the pumps that have succeeded. In other words, the target price mentioned in the pump message was successfully met by the market. The negatives of this task include all the negatives of Task I and some of the positives of Task 1---specifically, the pump messages that were not successful. In other words, the target price was not reached within 6 hours of the pump message. 

For both tasks, we train a binary \texttt{Random Forest} classifier. It might be possible to improve the classification accuracy by using sophisticated approaches, i.e., neural nets, but since our focus is to demonstrate the possibility of classification, we focus on traditional methods. We split the dataset into train/test, such that each sample $(c, t)$ corresponding to coin $c$ at time $t$ is in the training set if $t < d$ and in the test set otherwise. For each coin, $d$ is picked  in way to reach a $75\%/25\%$ split. 
Let $Train = \{S_1, S_2, ..., S_d\}$ and $Test = \{S_{d+1}, S_{k+2}, ... S_n\}$, sorted by their timestamps. For each coin, we train a separate classifier in the following way: \\
\begin{algorithm}[H]
\caption{}
\begin{algorithmic}
\State \textbf{Input:}
\State 1: Coin c
\State 2: $Train = \{S_i | S_i = (c, t_i)\}_{i=1}^d$ , sorted by $t_i$
\State 3: $Test = \{S_i| S_i = (c, t_i)\}_{i=d+1}^n$ , sorted by $t_i$
\For{$S_i$ in $Test$}
\State model = $RandomForestClassifier.train(Train)$
\State $P_i = model.inference(S_i)$
\State Add $S_i$ to $Train$
\State Remove $S_i$ from $Test$
\State Add $P_i$ to $Prob$
\EndFor
\State return $Prob$
\end{algorithmic}
\end{algorithm}

\begin{figure}
\centering
\begin{subfigure}[b]{0.8\columnwidth}
\includegraphics[width=\textwidth]{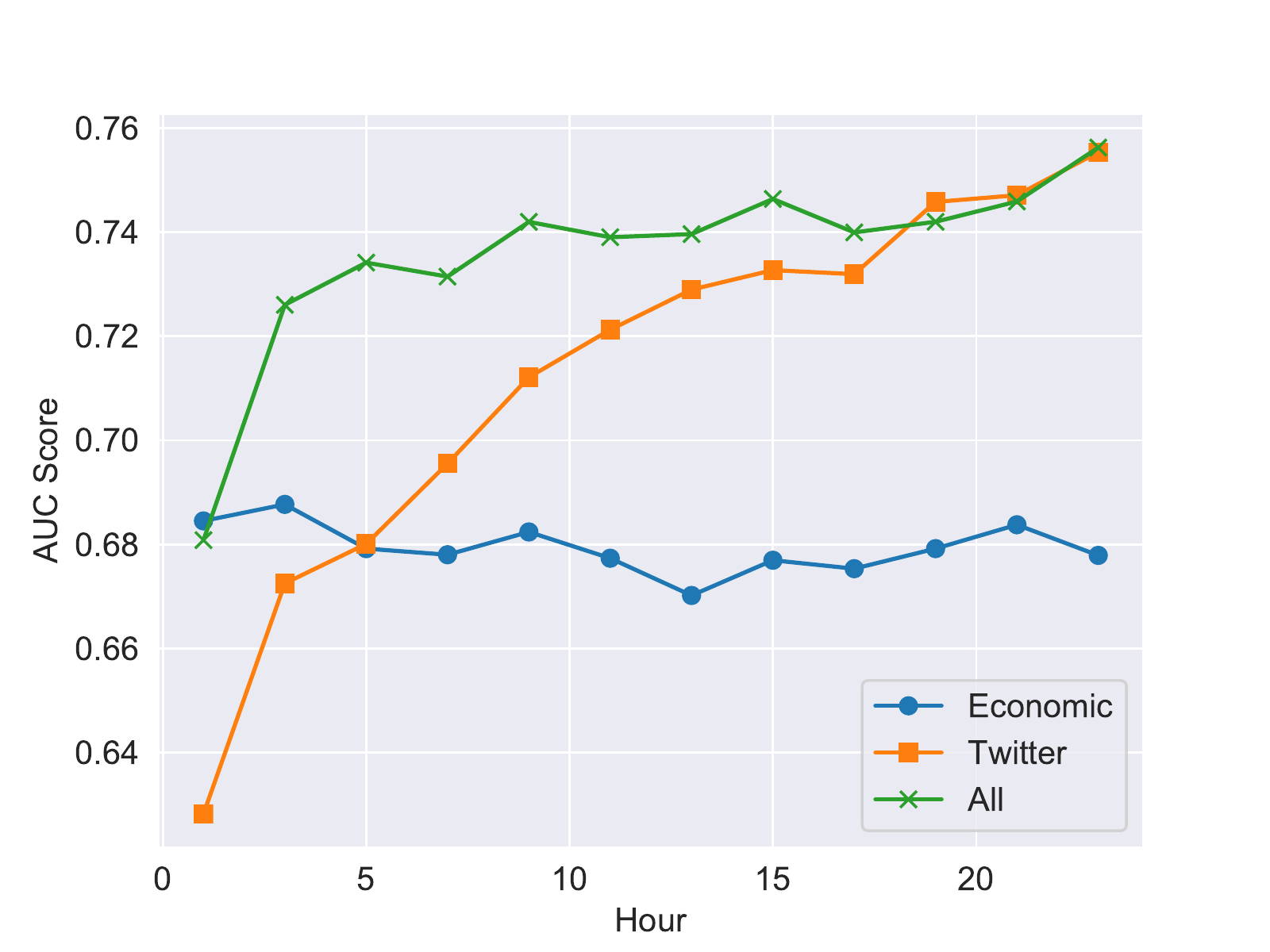}
\caption{Classification Task I}
\end{subfigure}
\quad
\begin{subfigure}[b]{0.8\columnwidth}
\includegraphics[width=\textwidth]{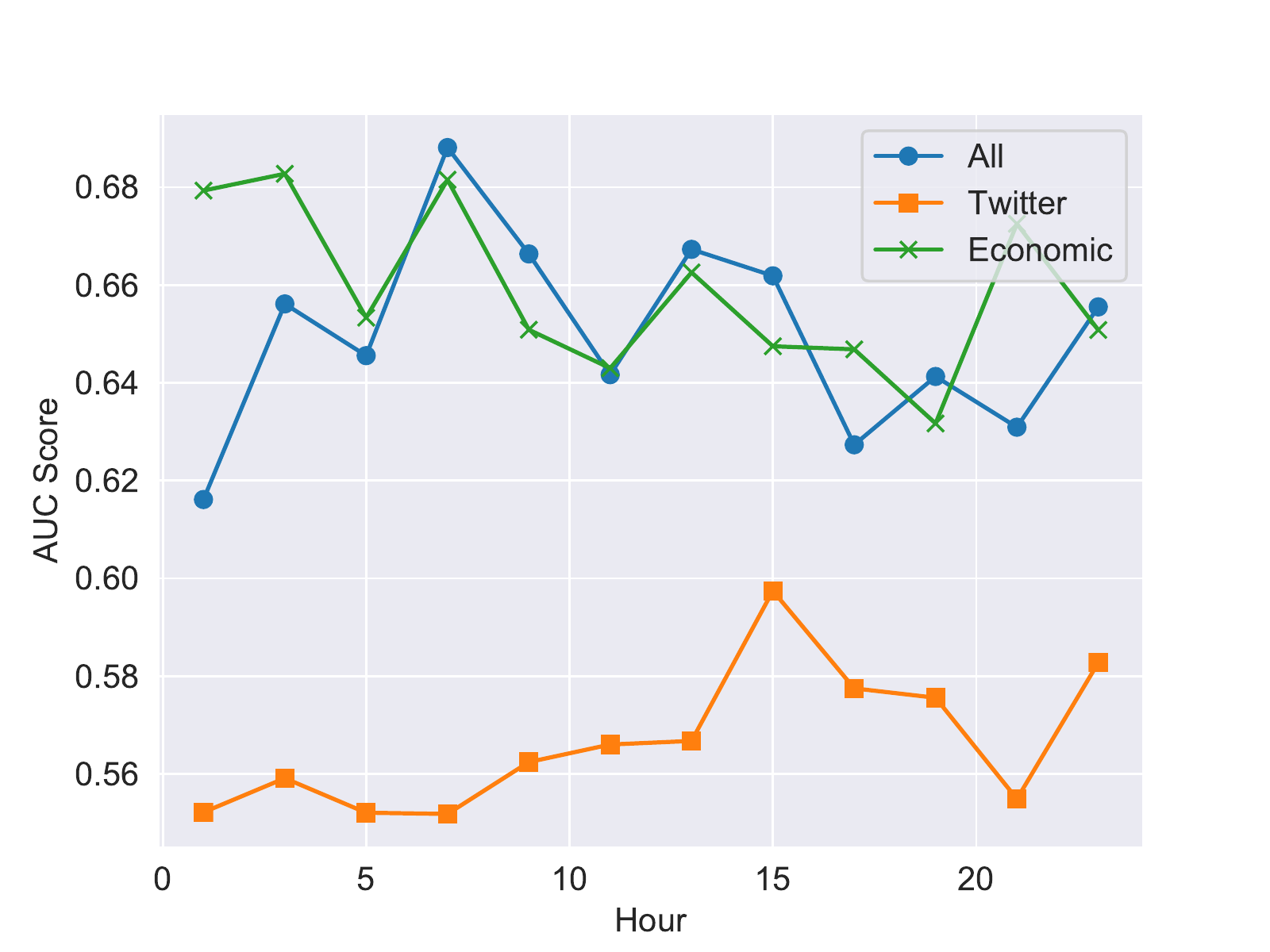}
\caption{Classification Task II}
\end{subfigure}
\caption {AUC-ROC score of the model, varying the time period of consideration from 1 to 24 hours.}
\label{fig:timelag}
\end{figure}

We evaluate the approach using the Area Under the Receiver Operating Characteristic Curve (ROC-AUC), which gives a baseline of $0.5$ for random guesses; the higher the metric, the better. The features are chosen as explained in Section \ref{sec:features}. Figure \ref{fig:timelag} shows AUC-ROC score by time period in the past. Based on this plot, for the first task, we choose $w=15$ for economic and Twitter features. For the second task, we choose $w=15$ for Twitter features and $w=7$ for economic features.

\subsection{Results and Discussion}
Table \ref{tab:exp1} summarizes Task I results. The columns describe the prediction accuracy measured in AUC score when using Twitter-only, economic-only, and combined features. When using a combination of features, the AUC score averaged over all the coins is $0.74$, which is significantly better than random baseline. We can see that social media features are more effective for this  task, although adding economic features provides a slight increase. Finally, when we take the average over only the top 20 highest volume coins, the prediction accuracy increases slightly. In the table we also show the AUC numbers for the five coins with the highest AUC score when using all the features.

\begin{figure}[t!]
\begin{subfigure}[t]{0.22\textwidth}
\includegraphics[width=\textwidth]{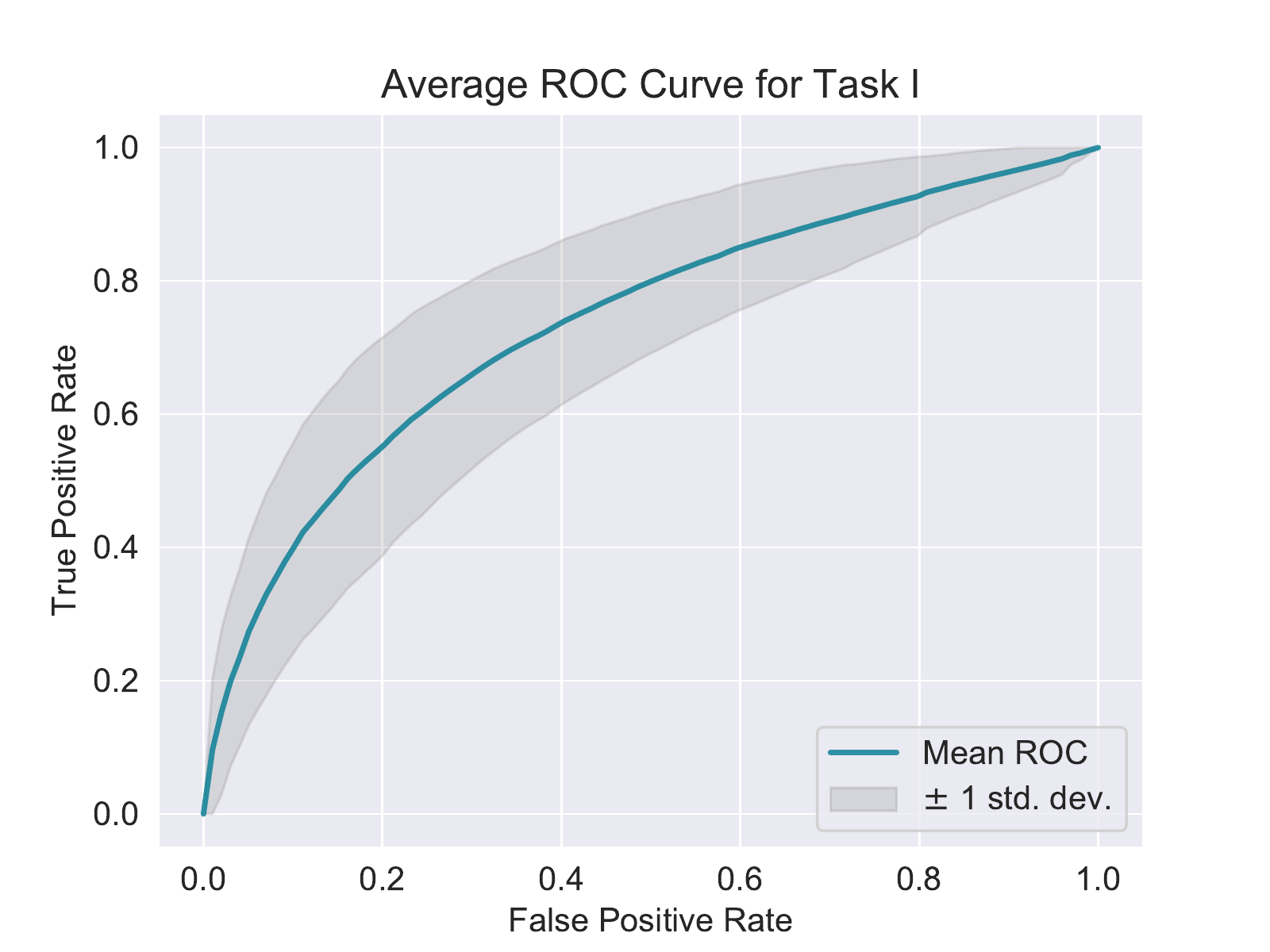}
\caption{Task I }
\end{subfigure}
\begin{subfigure}[t]{0.22\textwidth}
\includegraphics[width=\textwidth]{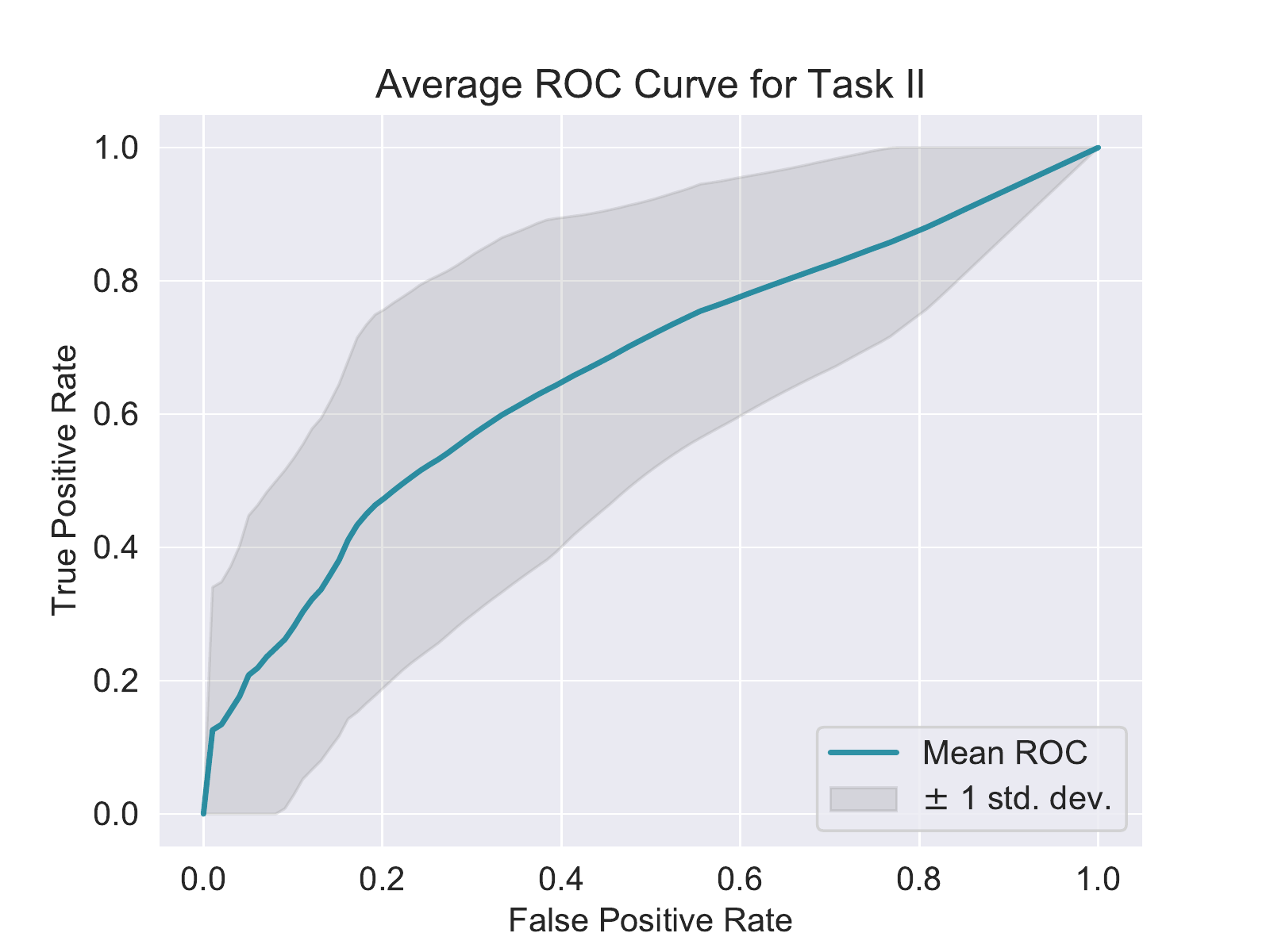}
\caption{Task II}
\end{subfigure}
\caption{Average ROC plot for the two classification Tasks that we consider. All features are used for the classification.}
\label{fig:roc}
\end{figure}

\begin{table}
\centering
\caption{ROC-AUC test performance on the two binary tasks, averaging 10 different train:test partitions. We show evaluations for three feature-sets. From right to left: The right-most is ``Both,'' utilizing both financial and Twitter. The ``Economic'' and ``Twitter'' utilize economic and Twitter features separately. The average AUC score is reported for (i) all the coins (ii) 20 coins with the highest dollar volume. The top five most-predictable coins are also shown for each task.}
\begin{subtable}{1\linewidth}
\centering
\caption{Classification Task I: Predicting Pump Attempts}


\begin{tabular}{l  c c c }
\toprule
 \textbf{} & \textbf{Twitter}&\textbf{Economic}&\textbf{Both}\\
 \midrule

$\textrm{Average AUC (all coins)}$ & $0.73 \pm 0.07$ & $0.67 \pm 0.1$ & $0.74 \pm 0.08$ \\
$\shortstack{\textrm{Average AUC (top 20 vol.)}}$ & $0.75 \pm 0.06$ & $0.68 \pm 0.1$ & $0.75 \pm 0.07$ \\
\midrule

\texttt{\$ADA} & $0.85 $ & $0.84$ & $0.88$ \\
\texttt{\$NCASH} & $0.84$ & $0.81$ & $0.88$ \\
\texttt{\$DGB} & $0.79 $ & $0.76$ & $0.86$ \\
\texttt{\$RCN} & $0.82$ & $0.61$ & $0.85$ \\
\texttt{\$TRX} & $0.83 $ & $0.81$ & $0.84$ \\

\bottomrule
\end{tabular}
\label{tab:exp1}
\end{subtable}

\begin{subtable}{1\linewidth}
\centering
\vspace{4mm}
\caption{Classification Task II: Will the Pump Succeed?}

\begin{tabular}{l  c c c }
\toprule
 & \textbf{Twitter}&\textbf{Economic}&\textbf{Both}\\
 \midrule
$\textrm{Average AUC (all coins)}$ & $0.59 \pm 0.16$ & $0.70 \pm 0.19$ & $0.66 \pm 0.17$ \\
$\shortstack{\textrm{Average AUC (top 20 vol.)}}$ & $0.62 \pm 0.18$ & $0.76 \pm 0.17$ & $0.71 \pm 0.16$ \\
\midrule

\texttt{\$NMR} & $0.71 $ & $0.36$ & $0.94$ \\
\texttt{\$XEM}  & $0.63$ & $0.66$ & $0.90$ \\
\texttt{\$XRP} & $0.77$ & $0.84$ & $0.90$ \\
\texttt{\$QTUM} & $0.83$ & $0.80$ & $0.89$ \\
\texttt{\$ARK} & $0.47$ & $0.77$ & $0.88$ \\
\bottomrule
\end{tabular}

\label{tab:exp2}
\end{subtable}
\end{table}

The results for Task II are shown in Table \ref{tab:exp2}. Overall, we see that the average accuracy is lower for this task compared to Task I, but is still considerably better compared to a random baseline. We also note that compared to Task I, the variance in the AUC scores is considerably higher. Interestingly, we observe that in average, economic features are much more useful for this task (average AUC of $0.7$) than Twitter features (average AUC of 0.59). Furthermore, adding Twitter features to the economic features actually deteriorates performance. The average ROC curve when using all features  is plotted in Figure \ref{fig:roc} for both classification tasks.


We suggest that the Twitter features are more predictive for the first task, due to the social nature of this process. To be more specific, when scammers target a coin (pump attempt), they try to promote it in different social media platforms, resulting in correlated activity on Telegram and Twitter. However, as we observed in Section \ref{sec:pump-efficacy}, only a small ratio of pump attempts ``succeed'' based on our definition. Therefore, although a pump operation might generate extra traffic in Twitter, its effect on the coin price depends on many other factors such as market characteristics.  This might be the reason that for the second task, using only market/economic features gives us better performance. One possible explanation for performance drop when we add Twitter features in the second task is that the number of positive labels for this task is very low, and adding Twitter features increases the sparsity.  Although we removed the coins with less than five positive in their training set, the number of positives is 14 in average for the remaining coins (less than 15\% ratio in average, and for some coins as low as 6\%). The average dimension of the Twitter features is 52 (it could vary across the coins, because of the number of the connected components) and more than a hundred for some coins. 

We investigated the reasons for obtaining high accuracy scores for some coins but not the others.  In particular, we analyzed potential relations between the prediction accuracy and financial indicators of a coin such as market cap, volume, and so on. As shown in Table~\ref{tab:exp2}, the accuracy score is typically higher for coins with higher dollar volume. Our preliminary analysis of other features produced rather ambiguous results, as we did not find meaningful correlations between accuracy and coin features. This may require more thorough investigation in the future.

\section{Prevalence of Twitter Bots}
\label{sec:bots}

In this section, we study the presence of bot activity around pump attempts by exploring the pump-user network that we explained in Section \ref{sec:corex-emb}. 
This bipartite network has 36 connected components, but the largest connected component contains roughly 99\% of the users and 99\% of pump attempts. Below we discuss the involvement of Twitter bots in those attempts. 

First, from our tweet dataset we extract the tweets containing a Telegram invitation link (e.g., http://t.me/Monsterpumper). We label the users associated with these tweets as \textbf{telegram active users.}

\begin{figure}
    \includegraphics[width=0.5\textwidth]{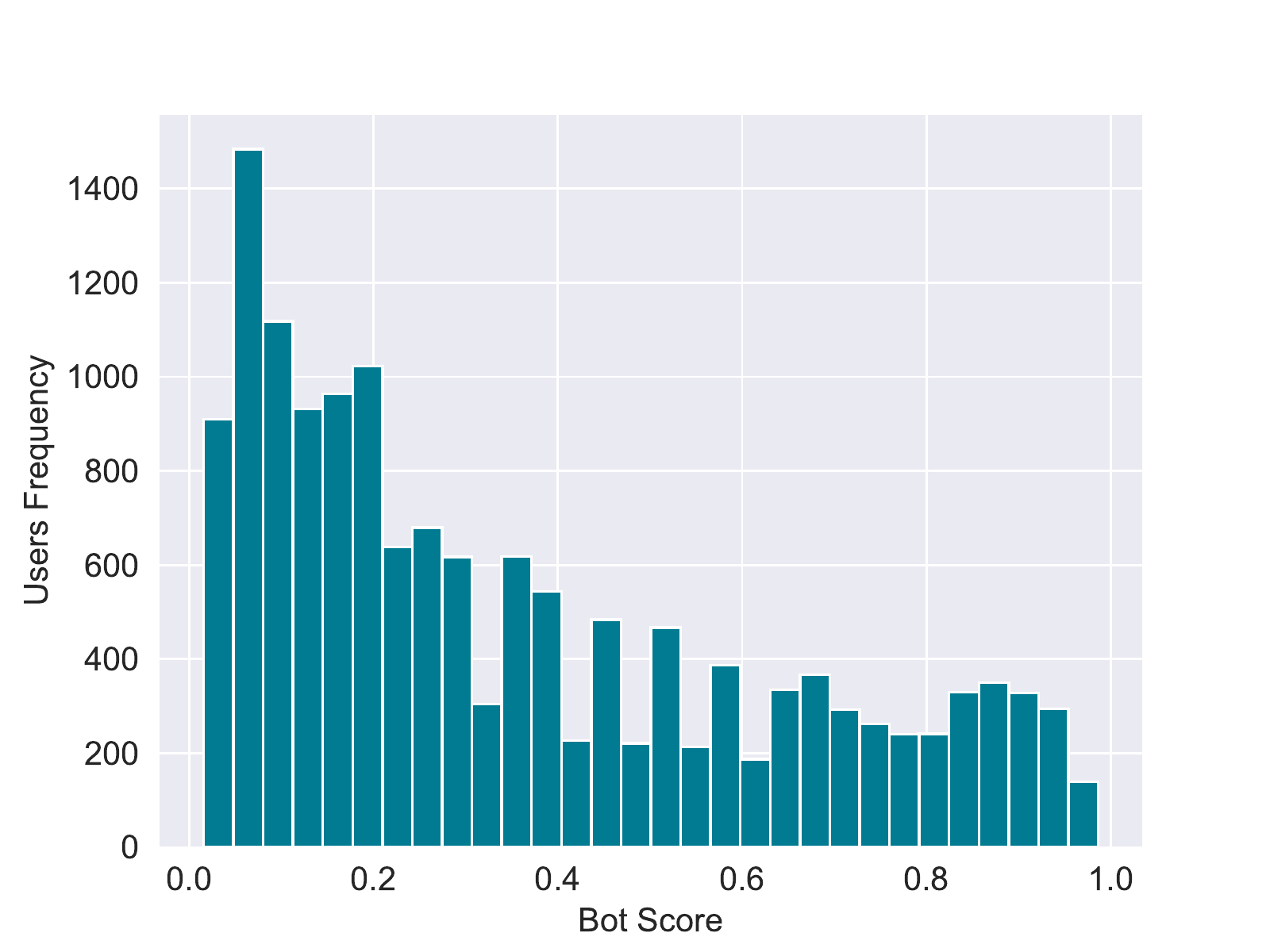}
  \caption{Score distribution obtained using Botometer API. Shows the probability of a user being bot.}
   \label{fig:english-bot}
\end{figure}

Next, we use two approaches for classifying a user as a {\em bot}. 
\begin{itemize}
    \item  Twitter Suspended List. Using the Twitter API, we collected the most recent account status of the users in our dataset and checked whether they are still \textbf{active} or \textbf{suspended} by Twitter. 
    \item Botometer Score \cite{varol2017online}. Twitter works based on algorithms with high precision but low recall, since they do not want to mistakenly suspend users that are not bots. So we employed the Botometer API  \footnote{https://botometer.iuni.iu.edu/\#!/api} to detect other potential bots. Given a user id, Botometer returns a probability of that user being a bot. Botometer classifies users using six types of features: friend, network, content, sentiment, temporal, and user. Figure \ref{fig:english-bot} shows the distribution of the classifier score using all six features.
\end{itemize}
A user was classified as a bot if either the user was suspended or its Botometer score is above $0.55$~\footnote{We use 0.55 instead of 0.5 suggested by ~\cite{varol2017online} to avoid edge cases}. 

Table \ref{tab:bot-stats} shows an increasing ratio by degree. For a given user, the degree is the sum of the weights of its adjacent edges in the pump-user network and is the number of times the user participated in the pump operations. This suggests that a larger fraction of highly active users are bots, and the ratio increases with the activity level. For example, $84\%$ of the users that participated more than 10K times in the pump activities are either suspended or are bots, according to their Botometer score. 

\begin{table}[t]
    \centering
    \caption{Ratio of scammer users based on the number of times a user contributes to the pump attempts. The ratio of Telegram active users and suspended users is higher among highly active users.}
    \label{tab:bot-stats}
    \begin{tabular}{c c c c c }
    \toprule
    \textbf{\shortstack{Degree}} & \textbf{\shortstack{Total\\ \# Users}}&\textbf{Suspended}&\textbf{\shortstack{Telegram \\ Active}} & \textbf{\shortstack{Botometer \\Score > 0.55}} \\
    \midrule
    50 & 20881 &  0.19 & 0.27 & 0.20 \\
    100 & 10969 &  0.23 & 0.34 & 0.20\\
    500 & 2577 & 0.40 & 0.37 & 0.24\\
    1000 & 1435 & 0.46 & 0.36 & 0.24\\
    5000 & 340 & 0.56 & 0.30 & 0.25\\
    10000 & 179 & 0.42 & 0.29 & 0.36\\
    \bottomrule
    \end{tabular}
\end{table}

\subsubsection*{User Clustering.} Now we look at different user clusters. Using the weight matrix $\mathbf{W}$ obtained from applying CorEx on B (explained in more detail in Section \ref{sec:corex-emb}) we cluster the users of the pump-user network by assigning user $u_i$ to $arg max \mathbf{W}_i$. 
 
\begin{figure}
     \centering
     \includegraphics[width=0.48\textwidth]{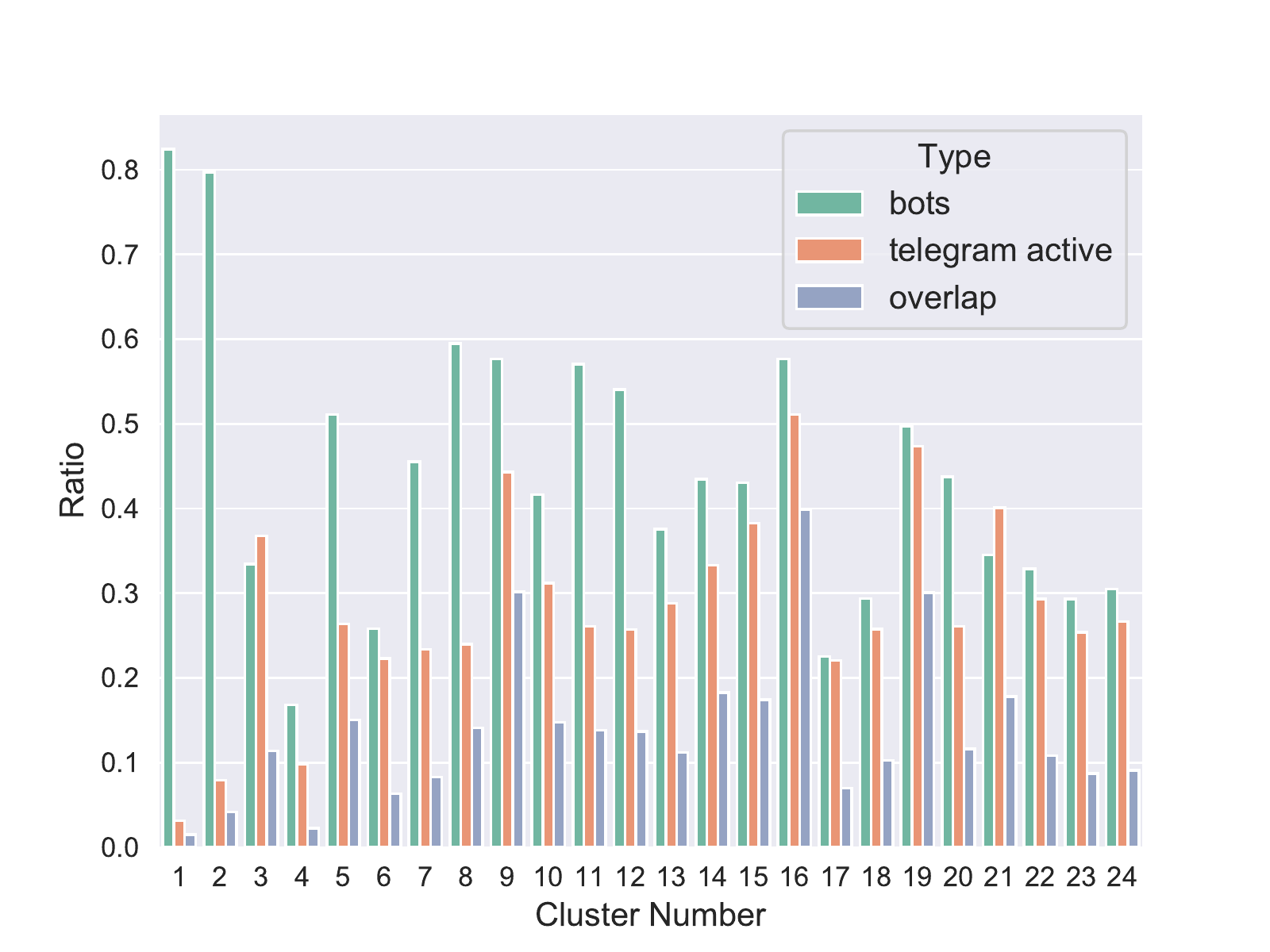}
     \caption{Ratio of bots and Telegram-active users in each cluster}
     \label{fig:bot-ratio}
 \end{figure}

Figure \ref{fig:bot-ratio} shows the ratio of the bots and Telegram-active users in each cluster. The blue bar shows the ratio of the users that are bots \textbf{and} Telegram-active. Note that the bot ratio is the number of users that are either suspended or have a botometer score > 0.55. The first and second cluster have 734 and 1,011 members respectively where more than 80\% of them are bots. Cluster 16 is also interesting in a sense that it has 600 members where 50\% of them are Telegram-active, and around 60\% are bots. Cluster 17 and 18 have 3K members each with a low ratio of bot members.
 
\section{Related Work}
We split the discussion of the related work into three complementary threads. First, we focus on fraud in social media and some of the efforts that have been taken to address it. Next, we discuss other work that analyzes the relation between financial market and social media data.  We then describe the effort on studying cryptocurrency activity, with a special focus on work that includes social media in its analysis.

Social media has a long history of fraudulent activity, and some types of fraud appear in our work. First, when scammers attempt to pump a coin by making it look more popular than it actually is, they are engaging in a specific type of misinformation. Misinformation is a major problem on social media~\cite{wu2016mining}, and several recent efforts have tried to detect it~\cite{starbird2014rumors,ratkiewicz2011detecting,kwon2013prominent}. Another burgeoning line of work is bot detection~\cite{darpabot}. There is a known connection between bots and misinformation, wherein bots are actively employed to spread misinformation in social networks~\cite{lazer2018science,forelle2015political}. We leverage previous literature in these areas in our approach. The bot labeling approach that we use in the bot assessment portion is based on previous work~\cite{hu2014social}. Additionally, we study the dynamics of users' reactions to the pump and dump campaigns. This is similar to previous work on social media where similar inputs are used to identify susceptible users~\cite{sampson2016leveraging,ozturk2015combating}.

The literature on predicting various financial market properties exploiting signals from different social media data is quite extensive. \cite{Chen2016AMood, Bellini2017ExploringAnalysis, Wang2017StockAnalysis} use sentiment features to predict stock price movements, while  \cite{Tu2016InvestmentMedia,Saumya2016PredictingNetwork,Ruiz2012CorrelatingActivity} employ network features extracted from various social media interactions. However, only a few papers study stock market manipulations on social media. \cite{renault2017market} studies pump and dump in OTC (Over The Counter or Penny stocks) and shows that an abnormally high number of messages on Twitter is associated with price increase followed by a price reversal. \cite{clarke2018fake} shows the presence of bot and spam activity on Twitter stock microblogs and compares Twitter activity with financial data. Authors in  \cite{cresci2019cashtag} uncover and introduce \textit{Cashtag Piggybacking}, a type of malicious activity on Twitter in which scammers try to highlight low-value stocks by co-mentioning them in the tweets with high-value stocks.

Several papers study cryptocurrencies, and analysts build models to predict their price movement. An example is~\cite{kim2016predicting}. In this paper the authors use cryptocurrency forums to predict the price and volume of cryptocurrencies. In another effort~\cite{phillips2017predicting} the authors build a model to predict price fluctuations of cryptocurrencies. Specifically, they use epidemic models on social media activity to predict price bubbles of cryptocurrencies.~\cite{kim2017bitcoin} further tests how users discuss cryptocurrencies and how that discussion impacts price. They found that specific topics are likely to be tied to price movements.~\cite{phillips2018cryptocurrency} extends this analysis by using wavelets to predict price movements based on social media data.~\cite{garcia2015social} looks into the dynamics underlying social media and how they correlate with cryptocurrency price. They find that opinion polarization has a significant effect on price, and use this to build a model that predicts the price of the cryptocurrency.
In an effort to understand the dynamics of cryptocurrency discussions,~\cite{linton2017dynamic} performed topic modeling on a popular cryptocurrency discussion forum. They identified several common threads of discussion, such as bitcoin theft. Moreover, they showed that different mining technologies have different patterns of adoption on the forums. 
Our work stands apart from these methods by moving away from predicting price and volume movements, and instead identifying patterns of malicious behavior.

The work that is most related to ours is  ~\cite{xu2018anatomy}, where the main goal is predicting which coin will be pumped based on social signals from Telegram. The authors focus on ``pre-pump'' messages that announce an upcoming pump operation, but do not mention a coin. They developed a model to predict the likelihood of each coin being the target of the subsequent pump operation following the ``pre-pump'' message. Our work is complementary in that we consider a richer set of prediction problems, we use social signals from Twitter, and we provide a user-centric analysis of such pump attacks.

\section{Conclusion}
In this paper we present a novel computational approach for identifying and characterizing cryptocurrency pump and dump operations that are carried out in social media. Specifically, given financial and Twitter data pertaining to a particular coin, our method is able to detect, with reasonable accuracy, whether there is an unfolding attack on that coin on Telegram, and whether or not
the resulting pump operation will succeed in terms of meeting the anticipated price targets. We also analyze activities of users involved in  pump operations, and observe a prevalence of Twitter bots in cryptocurrency-related tweets in close proximity to the attack. 





In future work, we plan to augment our datasets with other sources (e.g., Reddit posts) to help with the prediction tasks considered here. Also, while our analysis of bot activity relied on suspended accounts, it will be interesting to develop a bot detection tailored to the cryptocurency domain. Finally, as a practical outcome of the work presented here, we envision building a cryptocurrency monitoring system that will detect impending pump attacks in real-time and warn susceptible users.  

\section{Acknowledgements}
We thank Saurabh Birari for developing a crawler for collecting historical market data. We would also like to thank Emilio Ferrara and Pegah Jandaghi for their helpful comments.

\bibliography{pump-bib}
\bibliographystyle{ACM-Reference-Format.bst}

\end{document}